# Improved critical current densities in MgB$_2$ tapes with ZrB$_2$ doping

Xianping Zhang, Zhaoshun Gao, Dongliang Wang, Zhengguang Yu, Yanwei Ma [a)]

Applied Superconductivity Lab., Institute of Electrical Engineering, Chinese Academy of Sciences, P. O. Box 2703, Beijing 100080, China

S. Awaji, K. Watanabe

High Field Laboratory for Superconducting Materials, Institute for Materials Research, Tohoku University, Sendai 980-8577, Japan

**Abstract:**

MgB$_2$/Fe tapes with 2.5-15 at.% ZrB$_2$ additions were prepared through the *in situ* powder-in-tube method. Compared to the pure tape, a significant improvement in the in-field critical current density Jc was observed, most notably for 10 at.% doping, while the critical temperature decreased slightly. At 4.2 K, the transport Jc for the 10 at.% doped sample increased by more than an order of magnitude than the undoped one in magnetic fields above 9 T. Nanoscale segregates or defects caused by the ZrB$_2$ additions which act as effective flux pinning centers are proposed to be the main reason for the improved Jc field performance.

[a)] Author to whom correspondence should be addressed, E-mail: ywma@mail.iee.ac.cn

MgB$_2$ is considered as a possible substitute for Nb-Ti or high-T$_C$ oxide superconductors operated at around 20 K because of its high transition temperature (T$_C$) and weak-link free character. The low material costs, simple structure and small anisotropy are additional advantages of MgB$_2$. Therefore, it is thought as the most promising candidate for engineering application, especially for magnetic resonance imaging (MRI) magnet which prefers to work at cryogen-free circumstances. However, critical current density (J$_C$) of MgB$_2$ decreases rapidly under magnetic fields compared to those for the Nb-based superconductors. MgB$_2$ is a two-gap structure material, and phonon scattering induced by the lattice distortions in this material is very pronounced. This suggests that the upper critical field (H$_{C2}$) of MgB$_2$ can be enhanced by introducing electron-scattering defects. It has been demonstrated that C doped thin films have H$_{C2}$(0K) about 50 T [1]. Various dopants [2-8] have been investigated, leading to a wide variation in the reported values of H$_{irr}$ and J$_C$. In the case of ZrB$_2$, we reported earlier that the improved J$_C$ value was observed in 5% ZrB$_2$-doped MgB$_2$ tapes when sintered at 600°C [9]. Furthermore, an increase in H$_{C2}$ from 20.5 T to 28.6 T and enhancement of H$_{irr}$ from 16 T to 24 T were obtained in the ZrB$_2$ doped bulk MgB$_2$ samples as compared to the binary sample at 4.2 K [10]. This means that ZrB$_2$ doping is a very promising method for MgB$_2$ tapes to get higher J$_C$ in high magnetic fields. Therefore, it is very necessary to carry out further investigation on the doping effect of this material due to the lack of systematic study. In this work, a series of ZrB$_2$ doped MgB$_2$ tapes were prepared using an *in situ* powder-in-tube (PIT) method. The highest J$_C$ values were achieved in samples with 10% ZrB$_2$ doping level, more than 11-fold improvement compared to the undoped tapes. The field dependence of J$_C$ decreased by the ZrB$_2$ doping, suggesting that pinning centers effective in a high-field region were possibly introduced.

The detailed procedure for preparation of MgB$_2$/Fe tapes has been reported elsewhere [11]. Mg (325 mesh, 99.8%), B (amorphous, 99.99%), and ZrB$_2$ (2-5 μm, 99%) powders were used as the starting materials. Mg, B powders were mixed with the nominal composition of 1:2, the ZrB$_2$ doping levels were 2.5, 5, 10, 15 at.%, respectively. These mixed powders were packed into Fe tubes, then swaged, drawn and cold-rolled into tapes. The final size of the tapes was 3.2 mm width and 0.5 mm thickness. Undoped tapes were similarly fabricated for comparative study. These tapes were heated under an Ar atmosphere up to 800°C for 1 h, which was followed by a furnace cooling to room temperature.

The phase identification and crystal structure investigation were carried out using powder x-ray diffraction (XRD). Microstructure and composition analyses were performed using a scanning electron microscopy (SEM) equipped with an x-ray energy dispersion spectrum (EDX). DC magnetization measurements were performed with a superconducting quantum interference device (SQUID) magnetometer. The transport current $I_C$ at 4.2 K and its magnetic field dependence were evaluated at the High field laboratory for Superconducting Materials (HFLSM) at Sendai, by a standard four-probe technique, with a criterion of 1 $\mu$V cm$^{-1}$.

Figure 1 shows the XRD patterns for the series of $ZrB_2$ doped and undoped $MgB_2$ tapes. For the pure tapes, a small amount of MgO was detected as an impurity phase. We could not clearly see the existence of MgO from the XRD patterns of doped samples because of the superposition of the (220) peak for MgO and the (102) peak for $ZrB_2$. $ZrB_2$ phase can be identified in all the doped samples, and its peaks intensities increase prominently with increasing amount of $ZrB_2$ in the staring powder. For example, the (100) peak of $ZrB_2$ has the same intensity with (100) peak of $MgB_2$ at 2.5% doping level, but exceeds it with higher doping level. Contrast to Bhatia's experiments [10], a peak shift is not identified clearly in the XRD pattern for our tapes. This may be due to different methods used in XRD analysis. No other impurity phase was found from the XRD patterns, very similar to earlier reports [9, 10].

The relationships between critical temperature $T_C$ and the $ZrB_2$ doping level are summarized in figure 2. The $T_c$ onset for the undoped tapes is ~ 37.1 K. The $T_c$ decreased with increasing $ZrB_2$ doping level. However, $T_c$ has slightly dropped by 1.2 K for the 15 % high $ZrB_2$-doped tapes while around 0.5 K decrease in $T_c$ was observed in the 10% doped samples, which is in good agreement with the previous report [9]. On the other hand, all doping slightly depressed Tc (less than 1.2 K), indicating that the dopant incorporates into the $MgB_2$ structure. This may be caused by the ZrB$_2$ dispersion in the MgB$_2$ matrix, the effects of which are proposed to be the change in the electron diffusivities in the π and σ band [10]. However, the superconducting transitions widths were hardly changed with the 10% doping level, and became a little broader only at a 15% doping level. As we know, both ZrB$_2$ and MgB$_2$ have an AlB$_2$-type structure, and their lattice parameters are very similar, thus some of the dispersed ZrB$_2$ in MgB$_2$ matrix may be regarded as the defects of the MgB$_2$ grains, similar to the substitution of Mg by Zr in the Zr doped MgB$_2$ samples [12].

Figure 3 presents the J$_C$ in magnetic fields at 4.2 K for undoped and ZrB$_2$ doped samples. It is noted that the ZrB$_2$ doping significantly enhanced the J$_C$ values of MgB$_2$ tapes in magnetic fields. J$_C$ increased with the increase of ZrB$_2$ doping level, and reached the highest values at 10% doping level. At 4.2 K, the J$_C$ reached 6590 A/cm$^2$ at 9 T, more than 11-fold improvement compared to the undoped tapes. Then the J$_C$ decreased with further increasing the doping level (e.g. 15 %), which may be due to a large amount of ZrB$_2$ introduced. Although MgB$_2$ has relatively large coherence length, the existence of large amounts of impurity phases will bring weak links at grain boundaries [13]. On the other hand, the sensitivity of J$_C$ to magnetic fields was decreased by the ZrB$_2$ doping. Therefore, the ZrB$_2$ addition is supposed to introduce effective pinning centers in high field. This speculation is supported by the volume pinning force data plotted in figure 4, which clearly demonstrates an improved flux pinning ability by ZrB$_2$ doping. Comparing to the previous report [9], we found that the J$_C$ value of 5% doped MgB$_2$ tapes was much enhanced in the present work, in which a higher sintering temperature of 800°C was employed. Therefore, more segregates or defects could be introduced by ZrB$_2$ addition at higher temperatures, thus enhancing flux pinning and improving the high-field J$_C$.

Figure 5 shows the typical SEM images of the fractured core layers for undoped and ZrB$_2$ doped tapes. SEM results reveal that the MgB$_2$ core of the undoped samples was loose with some limited melted intergarin regions. In contrast, much larger melted regions of intergrains were observed in the ZrB$_2$-doped tapes, resulting in the better connectivity between the MgB$_2$ grains. This microstructural change is consistent with previous reports of other borides [14, 15] and Zr [12] additives. The grain boundaries may act as pinning centres in $MgB_2$ as in $Nb_3Sn$ [16], but the grain size for all our samples is almost the same (~0.2μm) observed from high magnification images (see figure 5 (b, d)), indicating that the enhanced $J_C$ of the $ZrB_2$ doped samples is not due to the grain-size difference. In addition, the good grain coupling mainly increases the $J_C$ values, hardly changes the field dependence of $J_C$. From the EDX mapping images (see figure 6) of the 10% $ZrB_2$ doped samples we can see that although there are some segregations of Zr, the Zr element was observed all over the $MgB_2$ core. Accordingly, it seems that $ZrB_2$ or Zr atoms formed solid solution with $MgB_2$ during the sintering process, thus introduces more point defects or nano-particles that can act as pinning centers. Therefore, the excellent Jc field performance is primarily due to nanoscale impurity precipitates or/and substituted crystal lattice defects introduced by $ZrB_2$ doping. The reduced $T_c$s of the $ZrB_2$-doped samples further supports this viewpoint.

All the doped tapes exhibited a superior field performance and higher $J_C$ values than the undoped tapes in a magnetic field up to 10 T, especially for 10 at.% doping level. The mechanism for the significant improvement of $J_C$-B performance may be explained by better connection between the grains and very strong pinning force in $ZrB_2$ doped samples. Feng et al reported that the addition of Zr element enhances the $J_C$-B characteristics of $MgB_2$ tape [12]. They concluded that the enhancement of $J_C$-B properties in high magnetic field is due to the reduction of grain size and small $ZrB_2$ particles formed by the Zr additive. In our experiment, there are no obvious grains size difference could be observed between the doped and undoped samples. By EDX mapping analyses we find that Zr element distributes all over the $MgB_2$ core. It is speculated that the distribution or solid solution of $ZrB_2$ or Zr with $MgB_2$ is the main reason for the enhancement of flux pining ability in our experiment. Moreover, from the shape of the $F_P$ profiles in figure 4 we can see that the relative pining force is more remarkable at 20 K, suggesting that further improvement of $J_C$–B performance is expected for the $ZrB_2$ doped samples at higher temperatures. As the main working temperature of $MgB_2$ is around 20 K, the $ZrB_2$ doping is a very favorable method for the fabrication of applicable $MgB_2$ tapes. It should be noted that the size of $ZrB_2$ particles used was 2-5 μm, the size of dopants will much larger than those in nano-particles dopes tapes [2, 8]. As the fine precipitates can work effectively as pinning centers, further $J_C$-B improvement is expected upon utilization of finer $ZrB_2$ particles.

In summary, we have studied the effect of $ZrB_2$ doping on the Jc-B properties of $MgB_2$ tapes prepared by an *in situ* PIT method. The phase compositions, microstructure, flux pinning behavior and transport property were investigated by x-ray diffraction, scanning electron microscopy, DC susceptibility measurements and transport measurements. It is found that the $J_C$ values have been significantly improved by $ZrB_2$ doping. The highest $J_C$ value was achieved for the 10 at.% doped samples. The enhanced field dependence of the $ZrB_2$ doped tapes is mainly due to more possible segregates or defects caused by $ZrB_2$ doping.

The authors thank Ling Xiao, Yulei Jiao, Xiaohang Li, Jiandong Guo, G. Nishijima and Liye Xiao for their help and useful discussion. This work is partially supported by the National Science Foundation of China under Grant Nos. 50472063 and 50377040 and National '973' Program (Grant No. 2006CB601004).

# Captions

Figure 1   XRD patterns of *in situ* processed undoped and all $ZrB_2$ doped tapes. The peaks of $MgB_2$ indexed, while the peaks of $MgO$ and $ZrB_2$ are marked by circles and asterisks, respectively.

Figure 2   Normalized magnetic susceptibility vs temperature for all the doped and undoped tapes.

Figure 3   Transport $J_C$-B properties at 4.2 K for undoped and $ZrB_2$ doped tapes. The measurements were performed in magnetic fields parallel to the tape surface at 4.2 K.

Figure 4   The normalized volume pinning force ($F_P/F_P^{max}$) versus magnetic field (T) at 5K and 20K for undoped and 10 % $ZrB_2$ doped tapes.

Figure 5   SEM images of the undoped (a, b) and 10% (c, d) $ZrB_2$ doped samples after peeling off the Fe sheath. The left and right columns show images with low and high magnifications, respectively.

Figure 6   EDX mapping images of the 10% $ZrB_2$ doped samples after peeling off the Fe sheath.

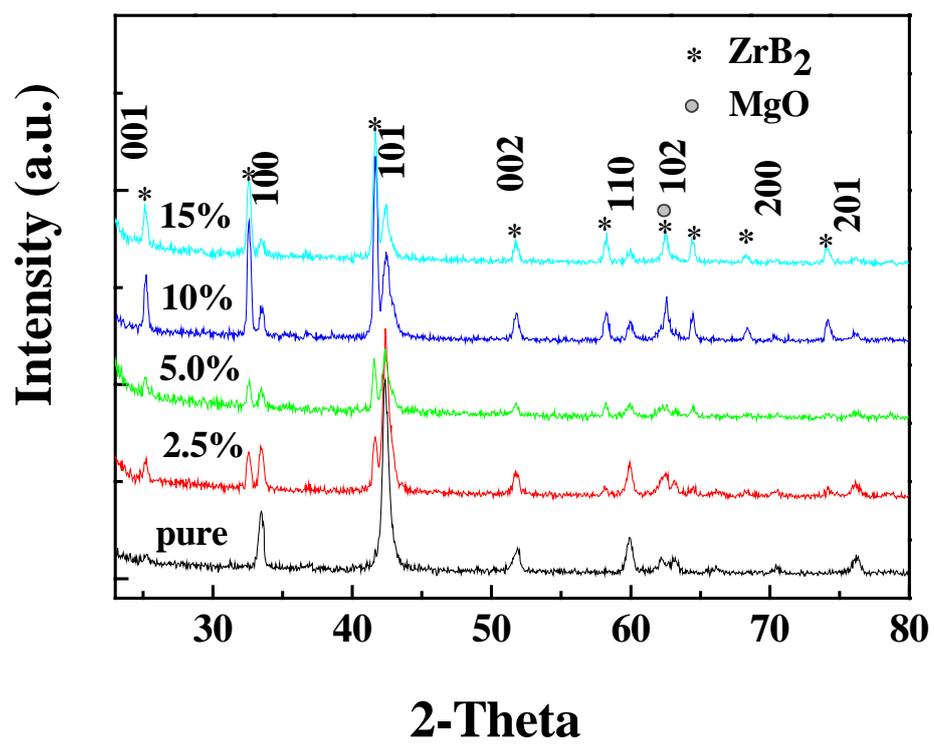

Fig.1 Zhang et al.

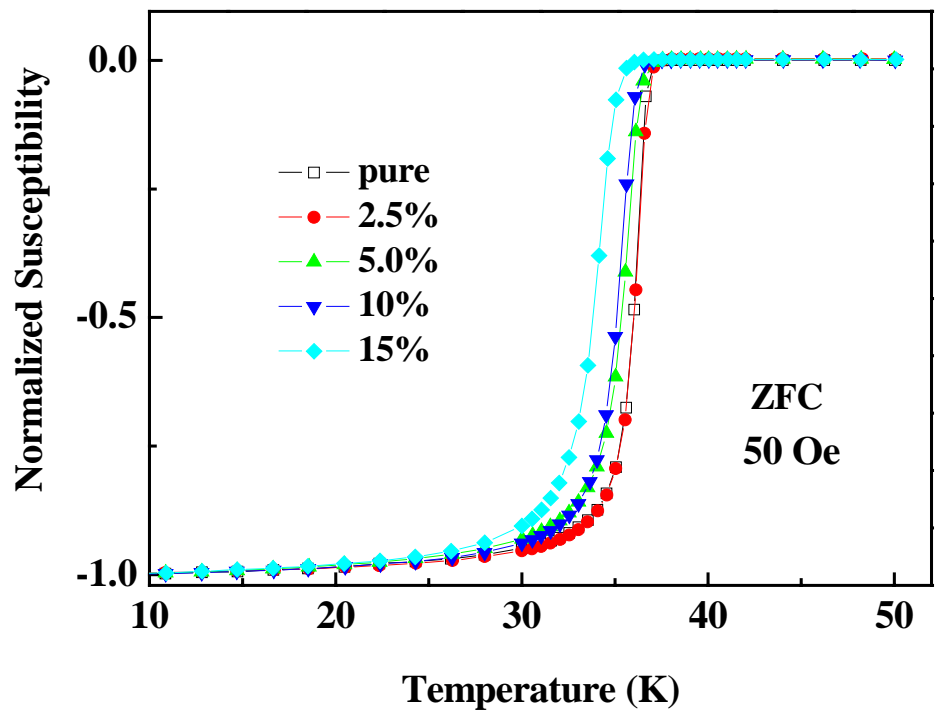

Fig.2 Zhang et al.

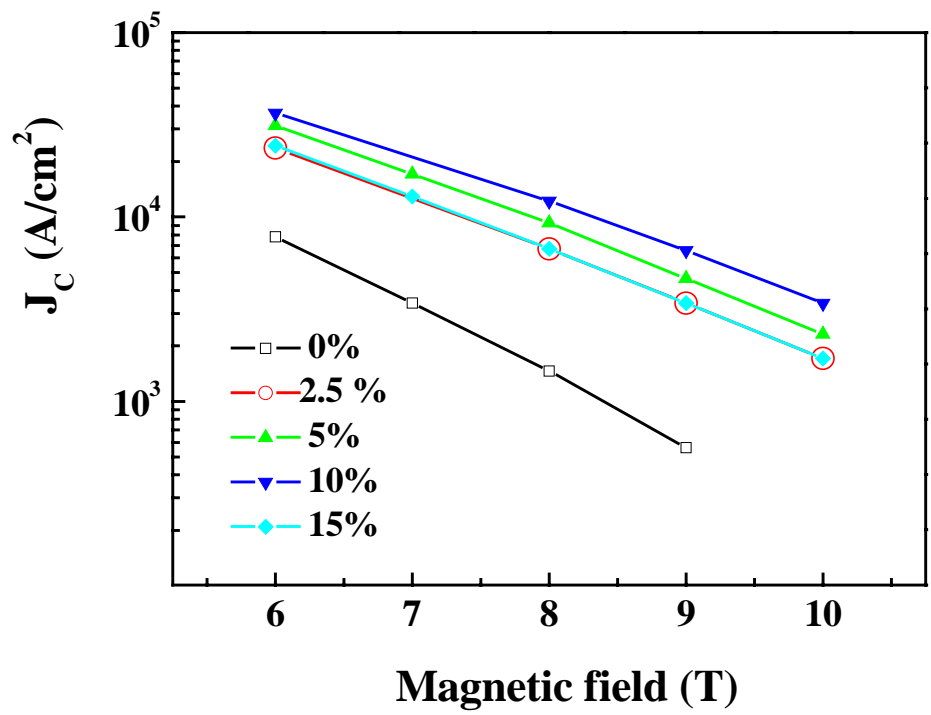

Fig.3 Zhang et al.

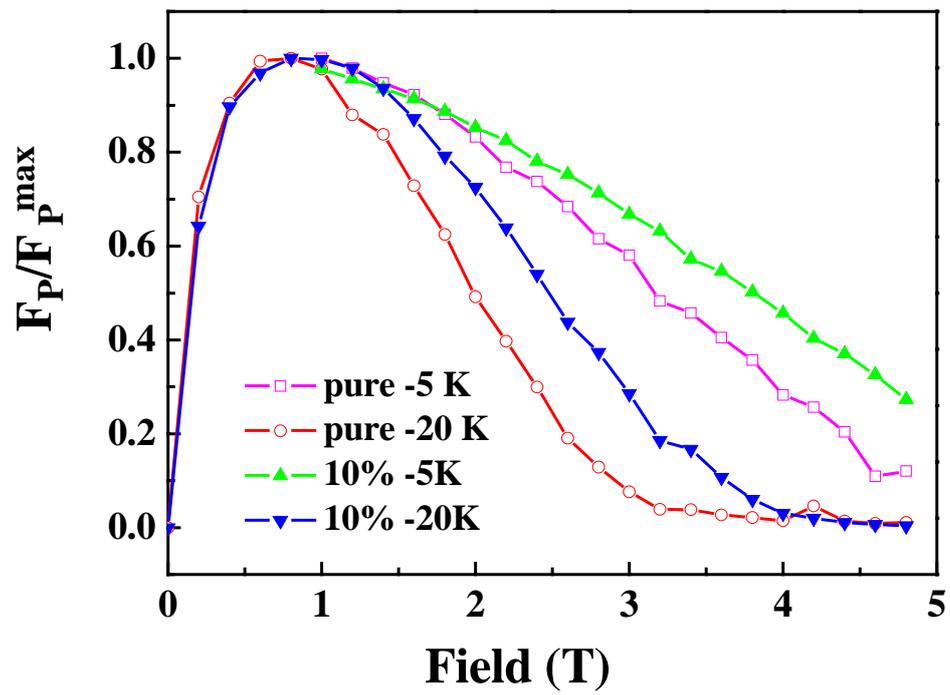

Fig.4 Zhang et al.

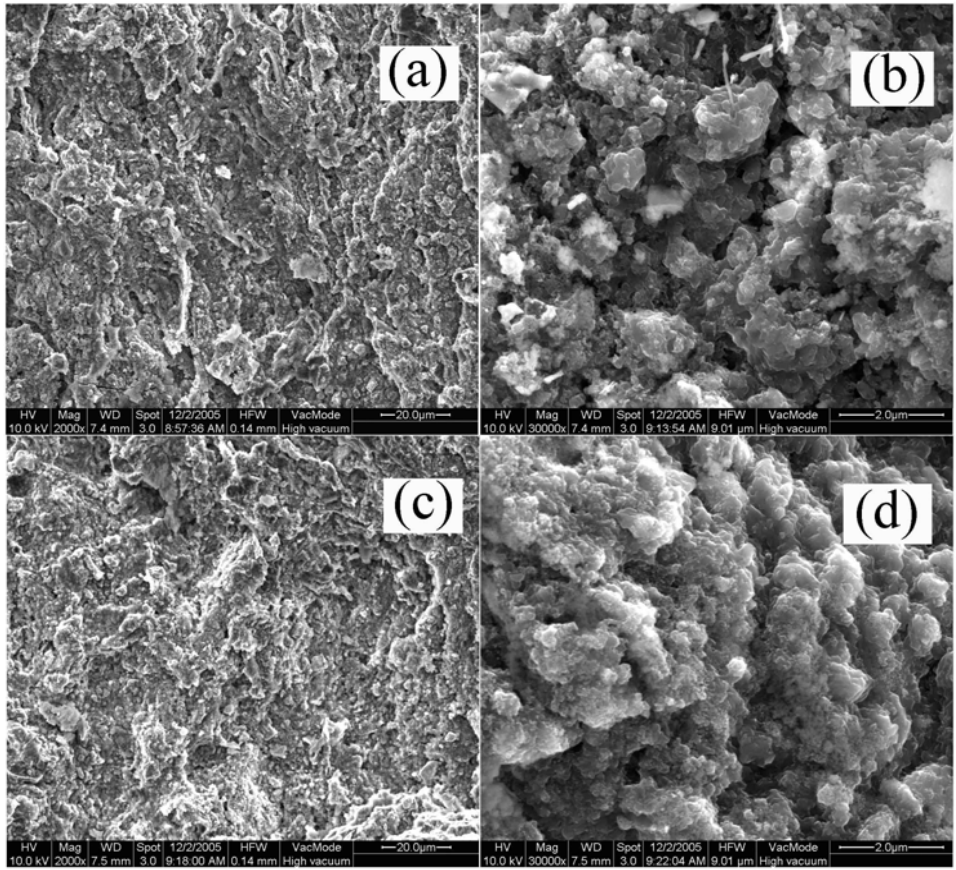

Fig.5 Zhang et al.

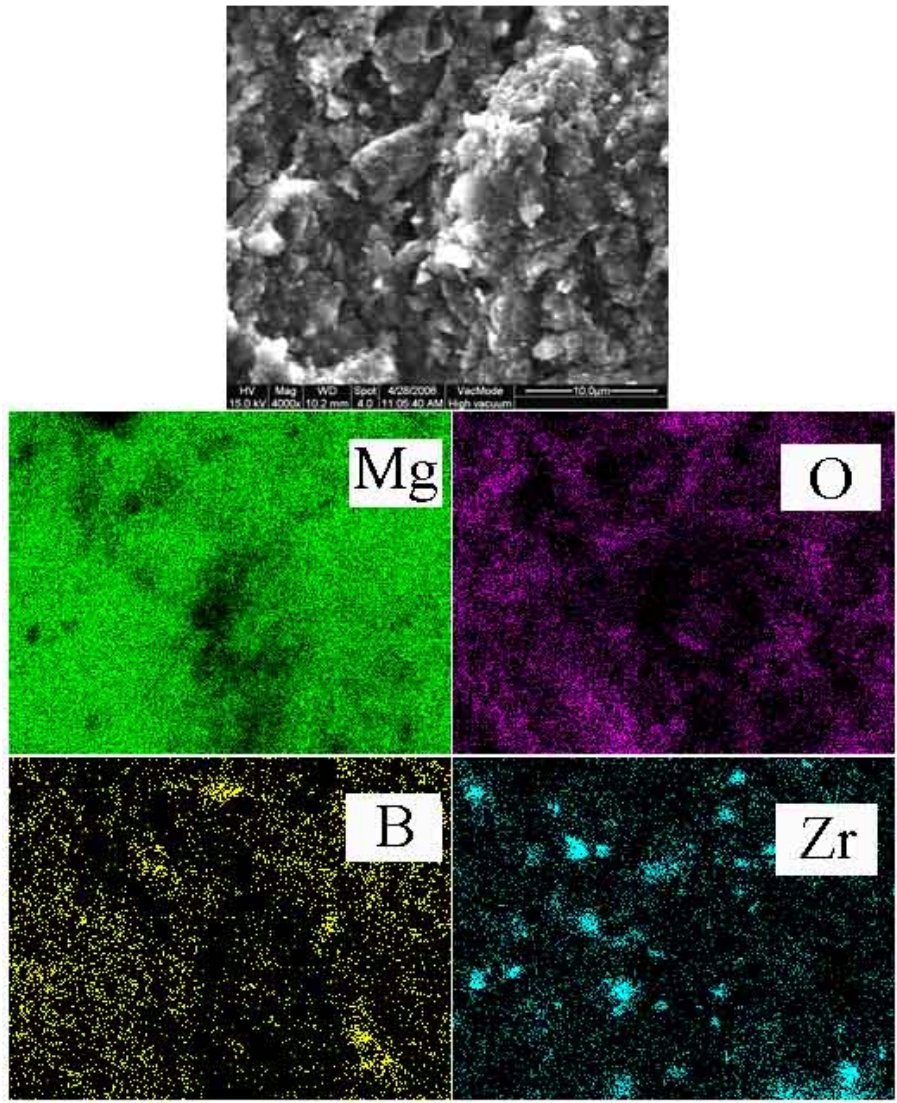

Fig.6 Zhang et al.